\documentclass[aps,12pt,final,notitlepage,oneside,onecolumn,nobibnotes,nofootinbib,superscriptaddress,noshowpacs,centertags]
{revtex4-2}

\usepackage{amsmath,amssymb}
\usepackage{mathtext}
\usepackage[utf8]{inputenc}
\usepackage{hyperref}
\usepackage{caption2}
\usepackage{graphicx}
\usepackage{multirow}

\makeatletter
\def\p@subsection  {}
\def\p@subsubsection  {}
\makeatother

\begin{document}

\title{Estimation of the capabilities of\\the SPHERE-3 Cherenkov telescope for determining\\the parameters of the primary cosmic particle}

\author{\firstname{E.\,A.}~\surname{Bonvech}}
\email{bonvech@yandex.ru}
\affiliation{Skobeltsyn Institute for Nuclear Physics, Lomonosov Moscow State University}
\author{\firstname{O.\,V.}~\surname{Cherkesova}}
\affiliation{Skobeltsyn Institute for Nuclear Physics, Lomonosov Moscow State University}
\affiliation{Department of Space Research, Lomonosov Moscow State University}
\author{\firstname{D.\,V.}~\surname{Chernov}}
\affiliation{Skobeltsyn Institute for Nuclear Physics, Lomonosov Moscow State University}
\author{\firstname{E.\,L.}~\surname{Entina}}
\affiliation{Skobeltsyn Institute for Nuclear Physics, Lomonosov Moscow State University}
\author{\firstname{V.\,I.}~\surname{Galkin}}
\affiliation{Skobeltsyn Institute for Nuclear Physics, Lomonosov Moscow State University}
\affiliation{Faculty of Physics, Lomonosov Moscow State University}
\author{\firstname{V.\,A.}~\surname{Ivanov}}
\affiliation{Skobeltsyn Institute for Nuclear Physics, Lomonosov Moscow State University}
\affiliation{Faculty of Physics, Lomonosov Moscow State University}
\author{\firstname{T.\,A.}~\surname{Kolodkin}}
\affiliation{Skobeltsyn Institute for Nuclear Physics, Lomonosov Moscow State University}
\affiliation{Faculty of Physics, Lomonosov Moscow State University}
\author{\firstname{N.\,O.}~\surname{Ovcharenko}}
\affiliation{Skobeltsyn Institute for Nuclear Physics, Lomonosov Moscow State University}
\affiliation{Faculty of Physics, Lomonosov Moscow State University}
\author{\firstname{D.\,A.}~\surname{Podgrudkov}}
\affiliation{Skobeltsyn Institute for Nuclear Physics, Lomonosov Moscow State University}
\affiliation{Faculty of Physics, Lomonosov Moscow State University}
\author{\firstname{T.\,M.}~\surname{Roganova}}
\affiliation{Skobeltsyn Institute for Nuclear Physics, Lomonosov Moscow State University}
\author{\firstname{M.\,D.}~\surname{Ziva}}
\affiliation{Skobeltsyn Institute for Nuclear Physics, Lomonosov Moscow State University}
\affiliation{Faculty of Computational Mathematics and Cybernetics, Lomonosov Moscow State University}


\begin{abstract}
New results of modeling the operation of the new SPHERE-3 Cherenkov telescope are presented. The telescope will be able to detect cosmic particles by direct and reflected Cherenkov light of the extensive air showers (EAS). Dual detection improves the accuracy of determining the parameters of the primary particle. The study is based on the data bank of distributions of the EAS Cherenkov light obtained on the Lomonosov-2 supercomputer. The accuracy of determining the energy and type of the primary particle from the reflected and direct flux of Cherenkov light is estimated.
\end{abstract}

\maketitle

{\let\thefootnote\relax\footnote{This a preprint of the Work submitted for publication in Physics of Atomic Nuclei, \copyright, copyright (2025), Pleiades Publishing, Ltd.}}\addtocounter{footnote}{-1}

\section{Introduction}

\begin{figure}[t!]
\setcaptionmargin{5mm}
\onelinecaptionstrue  
\includegraphics[width=0.5\linewidth]{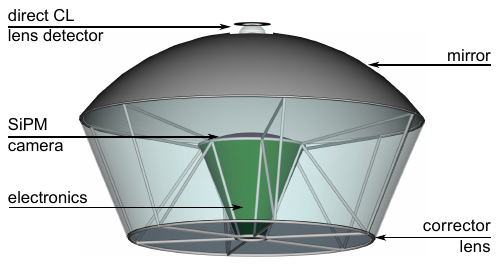}
\captionstyle{normal} \caption{Schematic representation of the SPHERE-3 detector.}
\label{fig:genview}
\end{figure}

In the SPHERE project, the study of cosmic rays with energies above 5~PeV is carried out by the A.E.~Chudakov method of registering the Cherenkov light of extensive air showers (EAS) reflected from the snowy surface of the Earth~\cite{Chudakov}. This method was successfully implemented in an experiment with the balloon Cherenkov telescope SPHERE-2~\cite{SPHERE-2}. Currently, a new telescope, SPHERE-3, is being developed, the main task of which will be to accurately study the chemical composition of primary cosmic rays with energies above 1~PeV. The telescope will rise above the Earth's surface to a height of 1.5~km using a special drone and register both the reflected light from the snow surface and the direct Cherenkov light from the same shower. The telescope will consist of two separate detectors, a detector of reflected Cherenkov light and a direct light detector. The general view of the detector design is given in~Fig.~\ref{fig:genview}. The design of the telescope's detectors will be optimized for the task of measuring the mass composition of primary cosmic rays. 

The proposed design for the reflected light telescope will incorporate a correction lens, in addition to the conventional Schmidt optical system, for this project~\cite{Chernov_PhAN_2024}. The design of the direct light telescope is not yet finalized, and several options are under consideration.

This article describes current methods and their accuracy for determining the parameters of the primary particle from reflected and direct Cherenkov light in the current model of the SPHERE-3 telescope. The peculiarities of determining these parameters related to the experimental features are shown.

\section{ORCHID Database}
The SPHERE-3 design is currently undergoing optimisation, and procedures are being developed to estimate parameters and create trigger algorithms for both detectors. In addition, a search is underway for the most sensitive characteristics of image shapes in the direct light detector. In order to develop methods for restoring primary parameters that are least sensitive to interactions and atmospheric models, a large database of events is required. This was accomplished using the powerful MSU Lomonosov-2 supercomputer~\cite{SC}. 

Extensive air showers were generated using the CORSIKA code~\cite{Heck1998}. The resulting output file recorded information about Cherenkov photons at four levels: the snow surface (455~m above sea level, i.e. the surface of Lake Baikal), and at possible telescope altitudes of 500~m, 1\,000~m and 1\,500~m above the snow surface. At the initial stage of calculations the maximal altitude was 2\,000~m instead of 1\,500~m. 

At the snow level, spatial and temporal information about the photoelectrons (i.e. those photons that will be detected by the telescope's mosaic detector and produce a photoelectron in the SiPM) is recorded. The spatial step is 2.5~m squared on a 3 km side, and the time resolution is 5~ns. 

At telescope flight altitudes (500~m, 1\,000~m, and 1\,500~m), the angular, spatial, and temporal distributions of Cherenkov photons are recorded. We started with the angular resolution of 1.0$^\circ$, but it turned out that this was not enough for the accuracy required to determine the characteristics of the particle based on direct Cherenkov light. Now the base angular resolution is 0.2$^\circ$. The spatial resolution is 10~m and the time structure comprises 13~bins, each 5~ns wide. 

We calculated the distributions of Cherenkov photons for six different primary nuclei (p, He, N, Al, S, and Fe) with energies of 5, 10 and 30 PeV using three different interaction models (QGSJET01~\cite{Kalmykov1993,Kalmykov1994, Kalmykov1997}, QGSJETII-04~\cite{Ostapchenko2011,Ostapchenko2014} and Sibyll~2.3~\cite{Gaisser1992,Gaisser1994,Engel1999,Gaisser2009,Riehn2015,Riehn2017,Riehn2020}). Furthermore, five different atmosphere models available in the CORSIKA package (1, 3, 4, 8, and 11) were considered. The zenith angles ranged from 5$^\circ$ to 30$^\circ$ in steps of 5$^\circ$, and the azimuth angle was chosen randomly. 

For each set of input parameters (energy, particle type, interaction model, atmosphere model, and zenith angle), 100 unique events were typically calculated. Currently, our event bank contains more than 100\,000 unique images of Cherenkov light distributions. We plan to extend our database to include events with 100~PeV energy and above. The statistics of the event number of the ORCHID (Open Refined Cherenkov Image Database) database are presented in table~\ref{database}. Researchers interested in our data can access them via our database on request.

\begin{table}[!h]
\setcaptionmargin{0mm}
\onelinecaptionstrue
\captionstyle{flushleft}
\caption{ORCHID database statistics \label{database}}
\bigskip
\begin{tabular}{|c|c|c|c|c|c|c|}
  \hline
    \phantom{XX}& \parbox[c][15mm]{25mm}{interaction\\model} & \parbox[c][15mm]{3cm}{nuclei} & \parbox[c][15mm]{25mm}{atmosphere\\models}  & \parbox[c][15mm]{15mm}{angle\\grid, $^\circ$} & \parbox[c][15mm]{25mm}{Altitudes, \\0.455 +, km } & \parbox[c][15mm]{15mm}{EAS\\count} \\    
  \hline
  1 & QGSJET01    & p, He, N, Al, S, Fe & 1, 3, 4, 11  & 1.0 & 0.5, 1.0, 2.0 & 27\,770\\
  2 & QGSJETII-04 & p, He, N, Al, S, Fe & 1, 3, 4, 11  & 1.0 & 0.5, 1.0, 2.0& 27\,224\\
  3 & QGSJETII-04 & p, He, N, Fe & 1, 3, 4, 8, 11  & 0.2 & 0.5, 1.0, 1.5 & 28\,800\\
  4 & Sibyll 2.3  & p, He, N, Fe & 1, 3, 4, 8, 11 & 0.2 & 0.5, 1.0, 1.5 & 36\,900 \\[1mm]
  \hline
\end{tabular}
\end{table}

Each unique simulated Cherenkov light distribution is used to create 100 copies of the original event by shifting the EAS axis spatially in parallel with the telescope's axis. These clones are then all processed through the telescope's optical system simulation, which is based on Geant4 toolkit~\cite{Agostinelli2003, Allison2006,Allison2016} and simulates the transport of photons through the detector of reflected or direct Cherenkov light. The final step is collecting photons on the detector camera. 

\section{Evaluation of EAS parameters}

The resulting light distributions in the detector of reflected Cherenkov light are fitted using a lateral distribution function, which enables the arrival direction, energy, and mass of the primary particle to be estimated.

The accuracy estimates for the determination of primary particle parameters for the reflected light detector, as described below, were obtained using a sample of 13,500 events, unless otherwise stated. This sample consisted of 1,500 events for each of the types of proton, nitrogen, and iron nucleus with energies of 5, 10, and 30~PeV.

\subsection{Reflected light detector model optimization \label{optimization} }

To validate the detector geometry, a dense carpet of photoelectrons was generated and directed towards the detector. In the simulation, a quarter of this carpet was projected onto the ground with a density of 50 photoelectrons per spatial array element. The analysis of the simulated data revealed that approximately 10\% of the light detected by the photodetector mosaic originated outside the nominal field of view. This effect is attributed to the presence of temporal shift lines in the signal, resulting from multiple and parasitic reflections within the optical system. Furthermore, it was observed that light entering from the designated sector was recorded across almost the entire mosaic, with up to 70\% of the mosaic segments responding to such events.

To improve the accuracy of the mathematical simulation of the optical response of the detector, all possible absorbers were incorporated into the detector design. In particular, an absorber was placed around each individual pixel to ensure that only light incident from above would be registered. The effect of these modifications was a more accurate reproduction of the carpet pattern in the detector mosaic and a significant reduction in the number of parasitic photoelectrons originating outside the detector’s field of view.

\begin{figure}
\setcaptionmargin{5mm}
\onelinecaptionstrue
\includegraphics[width=0.9\linewidth]{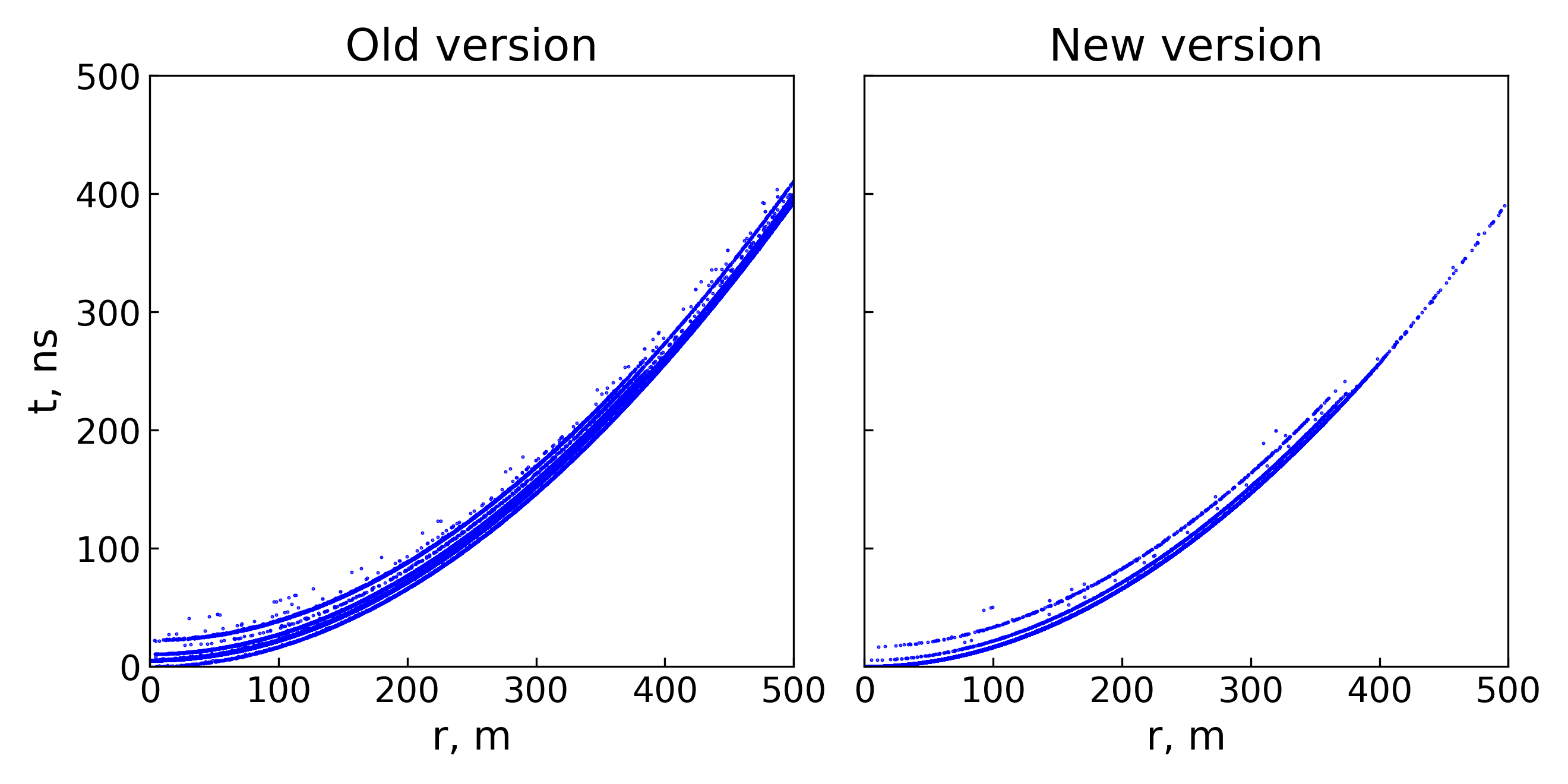}
\captionstyle{normal}
\caption{Distributions of optical path times, represented as individual points, for two detector geometries: the left panel corresponds to the previous (old) version, while the right panel illustrates the updated (new) configuration.\label{fig_1}}
\end{figure}

An analysis of the optical path times (see Fig.~\ref{fig_1}) of photoelectrons within the optical system revealed several persistent lines in the timing distribution, corresponding to parasitic photoelectrons and multiple refractions. The introduction of additional elements into the detector geometry led to a significant reduction in the number of such lines.

The modifications made to the detector geometry enabled the correct reproduction of the spatial pattern of photoelectrons on the detector mosaic and significantly reduced the contribution of light from outside the field of view. The number of parasitic photoelectrons and the intensity of temporal shift lines in the timing distribution were greatly reduced.

The effect of these improvements is demonstrated by comparing the spatial and temporal distributions of the signals in the previous and optimized detector geometries. In the updated configuration, the spatial pattern of the registered photoelectrons corresponds more accurately to the expected distribution, and the intensity of the temporal lines outside the detector field of view is significantly suppressed.

\subsection{Axis location \label{axis_location} }

\subsubsection{Selection of events within the field of view}

The position of the shower axis on the Earth's surface is determined by the distribution of light on the mosaic of photodetectors in the reflected Cherenkov light detector. The shower axis is defined as the projection of the center of mass of integrated signals in pixels around the maximum signal on the mosaic. Accuracy depends on the telescope height and is $5\pm3$~m for a telescope at 500~m and $7.5\pm4.5$~m for 1\,000~m height, if the axis falls within the field of view. If the shower axis is outside, the shower parameters cannot be restored. The main criteria for excluding events from the analysis are the position of the pixel with maximum signal at the edge of the mosaic. In some cases, e.g. if the shower is outside the detector field of view, this criterion may not be reliable. New criterions was developed as described below.

\subsubsection{Solution by classic methods}\label{sec:axis_standart}

The spatial distribution of the time-integrated signal in the mosaic of the reflected light detector is approximated by an axial symmetric function~\cite{Chernov_PhAN_2024} to reconstruct the energy and mass of a primary particle. This function is used in the criterion for determining the mass, and its integral serves as a measure of energy~\cite{Bonvech_super2025}. However, when the EAS axis is outside the telescope's field of view, significant information about the light distribution is lost. This leads to errors in determining the shape of the light distribution, and consequently, in the estimates of the primary parameters. The position of the axis is determined by the mosaic, even though it is actually outside it due to lack of information and fluctuations in the number of photons. When the number of detected photons is small, or when the axis is close to the edge, this can lead to significant errors.

A method for solving this problem is described in~\cite{Bonvech_super2025}. It was improved by increasing the sample and by optimizing the reflected light detector described in subsection~\ref{optimization}.

The idea of the method is to approximate the photon distribution by a plane. If the EAS axis lies outside the mosaic, only the peripheral part of the distribution reaches the detector. The shape of the light distribution on the mosaic becomes significantly flatter compared to events where the axis actually lies within the telescope's field of view. This difference can be used to construct a criterion based on the parameters available in the experiment that allows us to determine which axes are erroneously identified on the mosaic. Among several options considered, the criterion $K$ showed the best separation:

\begin{equation}
    K = \left(\frac{f_\text{surf}}{q_1} - \frac{f}{q_2} \right), 
    \label{plane}
\end{equation}

\noindent where, $f_\text{surf}$ and $f$ represent functions characterizing mean square ($\chi^2$) accuracy of the approximation by plane and axially symmetric functions, respectively, while $q_1$ and $q_2$ represent degrees of freedom for the approximation. The separation boundary based on the value of $K$ also depends on energy.

The criterion~\ref{plane} for the entire mosaic eliminated 53.1\% events with an incorrectly determined axis and 8.2\% those whose axis actually lies on it. For internal segments of the mosaic without two boundary layers of segments, it allowed removing 95\% of images with incorrectly determined axes, with a 8.6\% loss of correct data.

In the two boundary layers of the segments, the criterion~\eqref{plane} is not effective. However, for approximately 90\% of all events with an erroneously determined axis in the mosaic, the axis is determined on the last two mosaic segment layers. This led to the idea of introducing one another selection criterion: to exclude events in which the approximating function's axis falls on the last two segment layers of mosaic. In this case, approximately 30\% of the area is lost, but the reliability of determining the particle parameters increases.

The simultaneous operation of these two criteria makes it possible to eliminate 99\% of events where the axis lies erroneously on the mosaic, while losing 24--40\% of correct events, depending on the mass of the particle. This double selection criterion significantly improves the accuracy of energy estimation, as described in section~\ref{energy}.

\subsubsection{Solution by deep learning methods}
The development of a neural network-based method for filtering out events outside the field of view was based on simulated data. The simulated data included EAS events with different energies (5, 10, and 30~PeV) and primary particle types (protons, neutrons, and iron nuclei). The events were simulated at different zenith angles and at a fixed telescope altitude.

As a first approach, we implemented a convolutional neural network to predict the distance of the shower axis from the telescope. Based on this predicted distance, we determined the optimal threshold that allowed us to separate events with incorrectly identified exis. With a selected threshold of 210~m (which is slightly higher than the actual visibility limit of 180~m), we were able to filter out 71\% of the incorrect events with a loss of only 2.3\% of correct ones.

To further enhance the filtering performance, an additional technique based on an autoencoder with a spatial transformer network (STN) architecture was proposed. The STN allowed the neural network to independently determine the optimal orientation of the light spot, thereby improving the reconstruction of symmetric images that are representative of the true ball axes. Although the autoencoder alone does not possess strong filtering capabilities, its integration with the regression model allowed for a 85\% increase in filtering efficiency while maintaining a 97.8\% accuracy in detecting events with correctly located axis.

\subsection{Energy estimation}\label{energy}

Information about the position of the axis on the snow surface, and a full integral of an axially symmetric approximation function, is required to estimate the energy of a primary particle. For each energy, a set of dependencies is constructed for this integral with respect to the distance from the telescope's axis to the shower axis for every particle separately and for all particles, regardless of their type. The energy of the particle can be estimated by comparing the integral and distance to axes obtained experimentally with the set of these dependencies. If the particle's type is known, its energy can be determined using a set of dependencies for the particle. If the particle type is not known, then the energy is determined by a set of dependencies for all particles.


Errors were averaged over energy and mass. The accuracy of the energy estimation before applying the criterion described in section~\ref{sec:axis_standart} was 34\% for unknown masses and 27\% for known masses. After applying a criterion to reject incorrectly identified axes in the mosaic, the mean error decreased to 15\% and 8\%, respectively.


\subsection{Direction estimation}

The direction of the primary particle can be determined by reflected Cherenkov light and direct Cherenkov light. In case of a reflected Cherenkov light detector, the particle direction is estimated by constructing a projection of the time structure of the signal recorded in the mosaic, onto the snow, and then approximating it with a quadratic function~\cite{Chernov_PhAN_2024}. The precision in determining the angle of arrival of the primary particle using this method is slightly dependent on altitude and is between 1 and 2 degrees.

\subsubsection{Direction estimation by direct light}

The technique for determining the particle arrival direction by direct Cherenkov light was developed using both angular distributions of light that directly reaches the detector and images of Cherenkov light in the detector. The first option demonstrates the best accuracy that can be achieved with the chosen technique. The second method illustrates the accuracy expected in a real-world experiment with the detector.

\begin{figure}
\setcaptionmargin{5mm}
\onelinecaptionstrue
\includegraphics[width=0.9\linewidth]{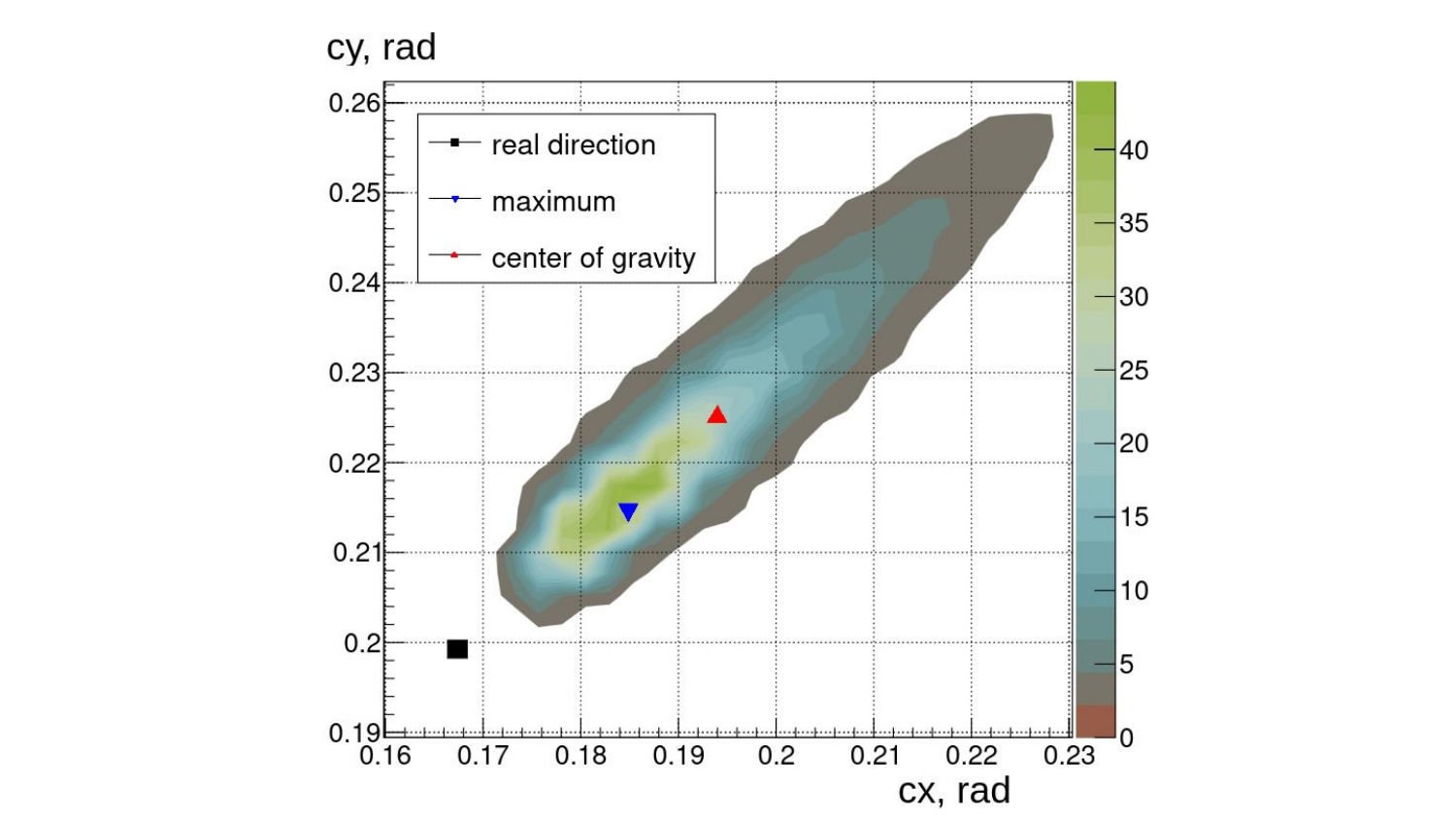}
\captionstyle{normal}
\caption{Angular distribution of Cherenkov light. The dots indicate the direction of arrival of the shower (square), the maximum (triangle up) and the center of gravity of the distribution (triangle down).\label{direct_image}}
\end{figure}

In both cases, the asymmetry of the light spot in the detector's field of view is used. This asymmetry lies in the fact that the maximum of the light distribution is far enough away from its center of gravity (see Fig.~\ref{direct_image}). A straight line drawn through the maximum and the center of gravity points to the real direction of the EAS axis. The difference between the direction of arrival and the center of gravity or maximum depends on the distance between the detector and the shower axis and is approximately 1--3$^\circ$. To account for this difference, we calculate its average value from model data for each distance (separately for maximum and center of gravity) and then subtract it from the coordinates of the corresponding point. The result differs from the actual direction by an average of 0.05$^\circ$ for the maximum and 0.14$^\circ$ for the center of mass at 100 m for the angular distribution. In the case of Cherenkov image analysis in a detector, the errors increase to 0.12$^\circ$ at the maximum and 0.16$^\circ$ at the center of mass, assuming a known axis position 100~m from the detector for different primary particles with an energy of 10~PeV.

\subsection{Particle classification}

\subsubsection{Classification by reflected light}

To evaluate detector geometry quality, we use criteria focused on the accuracy of PCR mass determination from the detector mosaic image, which is the main objective of the SPHERE-3 experiment. A method for estimating the mass composition of cosmic rays in the 1--100 PeV range for SPHERE--type detectors has been developed and implemented in software~\cite{Latypova2023_Zapiski}. This work employs the latest version of the Cherenkov photon lateral distribution function approximation, described in detail in~\cite{Latypova2023_Zapiski}. Here, a one-dimensional criterion based on the LDF projection is used to differentiate the mass of the primary particles.

The shape of the Cherenkov light distribution, characterized by a dimensionless parameter calculated from the measured image, strongly correlates with the longitudinal development of the shower. The criterion is designed to be integral, accounting for most of the light in the image and suppressing fluctuations, and only weakly dependent on hadronic interaction models, as it relies on relative, rather than absolute, distribution parameters~\cite{Chernov_PhAN_2024}. 

For showers at a zenith angle of 10$^\circ$ and a primary energy of 10~PeV (proton, nitrogen, iron), the maximum separation error was selected as the main criterion. The proton-nitrogen boundary had a misclassification rate of 36.4\% for protons; the nitrogen-iron boundary had a misclassification rate of 36.7\% for iron showers.

\subsubsection{Classification by direct light}

The overall aim of the SPHERE project is to develop a method to ascribe mass values to the primary particles of the registered EAS. For the purpose we need a Cherenkov light distribution parameter sensitive to the mass. A simple and relatively cheap way to test a parameter is to construct a classification scheme of all the nuclei into 3 classes (protons, nitrogen, and iron).

At present, the primary parameter being considered is the length of the major axis in the Cherenkov image. For a proton-nitrogen classification, if the length of this axis exceeds a certain threshold value, the particle is categorized as a proton. Conversely, if the axis length is below the threshold, the particle is assigned to the nitrogen category.
In the case of a nitrogen-iron classification, particles with an axis length exceeding the threshold are classified as nitrogen. If the axis is shorter than the threshold, they are categorized as iron particles.

Since the length of the major axis of a Cherenkov light spot is a shape parameter that weakly depends on the nuclear interaction model, it also depends on other characteristics of the shower, such as the energy of the primary particle, the distance of the detector from the shower axis at flight altitude, and its azimuthal position relative to the shower axis. Therefore, it has become necessary to introduce a grid of criteria based on the azimuthal position and distance from the detector. Without using this grid, the probability of misclassification of the primary particle could reach 0.4.

To further improve the criteria, it was necessary to employ an absolute threshold that depends on distance. This threshold is the number of photons in each pixel of the Cherenkov light image. When this number exceeds a certain value, the pixel is considered to be part of the image. Due to the increased amount of light near the shower axis, as the detector gets closer to it this value increases. Using a distance-based threshold, we were able to achieve classification error rates between 0.23 and 0.31.

\subsubsection{Double mass classification}

Since SPHERE-3 will include both direct and indirect light detectors, a method for estimating mass using data from both types of detectors has been developed. Since the data are independent, this approach will reduce classification errors, as will be demonstrated later.

\begin{figure}
\setcaptionmargin{5mm}
\onelinecaptionstrue
\includegraphics[width=0.7\linewidth]{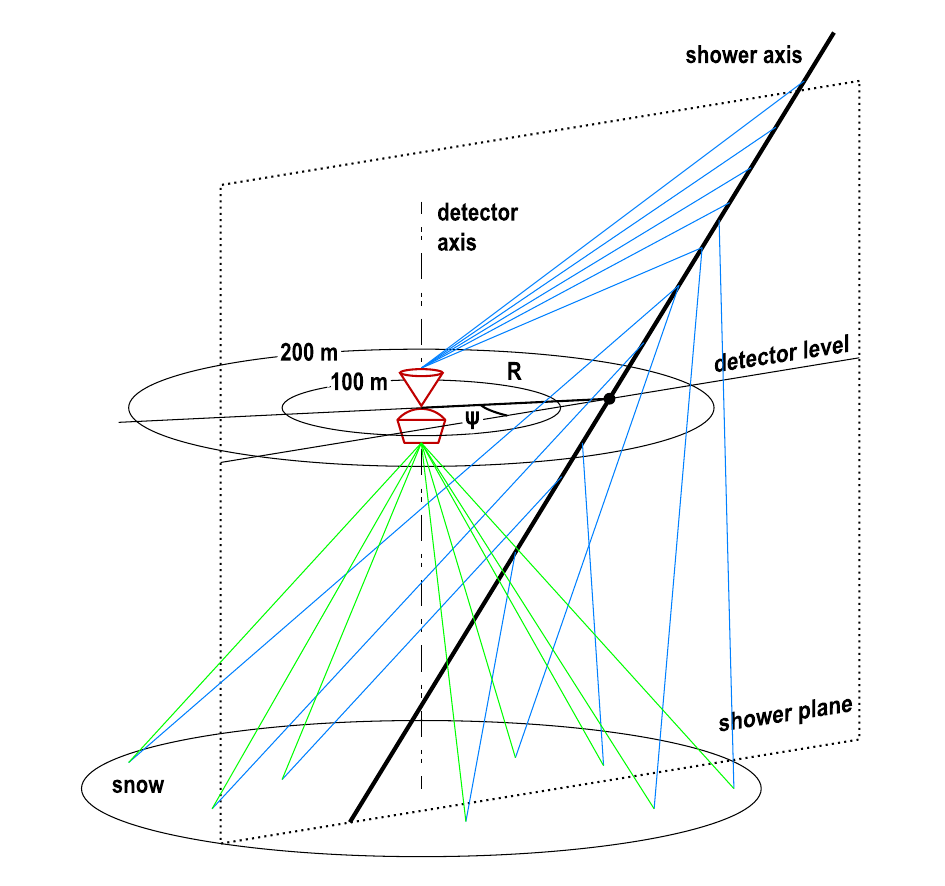}
\captionstyle{normal}
\caption{EAS detection at two levels geometry. Reflected light (green) is collected by the reflected light detector of the SPHERE-3, direct light (blue) is accepted by the upper one.}
\label{fig:dual}
\end{figure}

It should be noted that in order to implement a dual classification at an altitude of 500~m above snow level, the distance from the detector to the shower core must be between 100 and 200~m at flight altitude, and the distance between the intersection of the shower core with the snow surface and the detector's projection on the snow must be less than 175~m. Modelling has indicated that approximately 1/3 of such events occur. The relative positions of the detectors and showers during dual registration are illustrated in Fig.~\ref{fig:dual}.

For the purpose of the double classification, each suitable event was assigned a point in a two-dimensional feature space. For direct light, the length of the major axis was utilized, while for reflected light, the ratio of integrals along the inner and outer circles of the image was used. Subsequently, by sorting coefficients, a line that minimizes classification errors was determined.

All the improvements mentioned above were applied to direct light. The use of double classification reduced the classification errors to between 0.14 and 0.22.

\section{Conclusion}
The current findings regarding the determination of the primary particle's parameters are presented in this paper. The results of separate procedures designed to independently determine the particle parameters are presented. Both independently and jointly, particles are classified according to their mass based on the data from the two detectors. Detecting a particle by both detectors increases the accuracy in determining its type when compared to using each detector individually. In the near future, it will be necessary to develop a self-consistent process for determining all particle parameters, taking into account data from both detectors.

\begin{acknowledgments}
The research was carried out using equipment of the shared research facilities of HPC computing resources at Lomonosov Moscow State University~\cite{SC}.

This work is supported by the Russian Science Foundation under Grant No. 23-72-00006, https://rscf.ru/project/23-72-00006/

\end{acknowledgments}

\end{document}